# Los beneficios de un sistema CRIS para la investigación científica

Kevin Lajpop y Ana Ixcolin
*Centro de Investigaciones de Ingeniería, Universidad de San Carlos de Guatemala*
*CII-USAC*
*Guatemala, Depto. de Guat.; Guatemala*
[kalajpop20, aixcolin]@ingenieria.usac.edu.gt

*Abstract*— The visibility of research projects is crucial for maximizing their scientific and societal impact. Current Research Information Systems (CRIS) centralize data, enhancing access, the dissemination of results, and interdisciplinary collaboration. This article examines how CRIS improves global reach, funding opportunities, and adherence to open science principles, contrasting these advantages with the limitations faced by projects without such systems. CRIS proves to be essential tools for optimizing information management, strengthening research, and positioning institutions within the global scientific community.

*Resumen*—La visibilidad de los proyectos de investigación es clave para maximizar su impacto científico y social. Los sistemas de información de investigación (CRIS) centralizan datos, mejorando el acceso, la difusión de resultados y la colaboración interdisciplinaria. Este artículo analiza cómo los CRIS potencian el alcance global, las oportunidades de financiamiento y la adherencia a la ciencia abierta, contrastando con las limitaciones de los proyectos sin esta herramienta. Los CRIS demuestran ser esenciales para optimizar la gestión de información, fortalecer la investigación y posicionar a las instituciones en el ámbito científico global.

*Palabras claves*—CRIS, Ciencia, Investigación, Tecnología.

## I. Introducción

En el contexto actual de la investigación científica, la gestión, difusión y visibilidad de los proyectos de investigación han cobrado una importancia sin precedentes. Con el avance hacia un modelo de ciencia abierta y colaborativa, las instituciones de investigación enfrentan el reto de hacer accesibles sus resultados a un público más amplio, tanto dentro de la comunidad científica como hacia la sociedad en general. En este escenario, los sistemas de información de investigación, conocidos como CRIS (Current Research Information Systems), han emergido como herramientas clave para optimizar la organización, el acceso y la visibilidad de los proyectos de investigación.

Un sistema CRIS permite centralizar y gestionar de manera estructurada la información de cada proyecto, desde su etapa inicial hasta la publicación de sus resultados. Esta capacidad no solo incrementa la eficiencia en la administración de datos de investigación, sino que también fortalece el impacto de los proyectos al facilitar su difusión y permitir un acceso más amplio y transparente. Los sistemas CRIS ofrecen una infraestructura que permite a las instituciones académicas no solo gestionar sus investigaciones, sino también amplificar su alcance a nivel global [1].

En las instituciones que no cuentan con un sistema CRIS, la visibilidad de los proyectos suele estar fragmentada y sujeta a la dispersión de datos en múltiples plataformas, lo que limita la capacidad de los investigadores para difundir sus resultados y conectar con otros científicos interesados en colaboraciones. Este artículo explora cómo un sistema CRIS puede transformar la visibilidad de un proyecto de investigación, proporcionando un marco de referencia que permite entender las ventajas de su implementación en términos de acceso, difusión, colaboración y apoyo al modelo de ciencia abierta.

A lo largo del artículo, se analizan diversos aspectos de la visibilidad de los proyectos de investigación con y sin un sistema CRIS, y se presenta un análisis cuantitativo para evaluar el impacto de estos sistemas en la difusión y alcance de los proyectos científicos. Además, se discuten las implicaciones del uso de sistemas CRIS en la integración con redes de investigación y en la obtención de financiamiento, aspectos cruciales para el desarrollo científico en un entorno cada vez más competitivo y orientado a la colaboración.

Con esta investigación, se busca proporcionar una visión integral sobre los beneficios de los sistemas CRIS para la investigación científica, subrayando su papel en la promoción de un entorno de investigación más accesible, estructurado y alineado con las tendencias actuales de la ciencia abierta. Este análisis pretende servir como base para que instituciones académicas y de investigación consideren la adopción de un sistema CRIS como una estrategia esencial para potenciar la visibilidad y el impacto de sus proyectos.

## II. Funcionalidad y beneficios clave de los sistemas CRIS

Un Sistema de Información de la Investigación Actual (CRIS, por sus siglas en inglés) es una plataforma tecnológica diseñada para centralizar y gestionar de manera eficiente toda la información relacionada con proyectos de investigación. Este tipo de sistema tiene como objetivo organizar datos sobre investigadores, publicaciones, proyectos, financiamientos y resultados en un único lugar, facilitando su acceso y análisis tanto para los investigadores como para las instituciones educativas [10].

Los sistemas CRIS se destacan por su capacidad de integrar y automatizar procesos que tradicionalmente eran gestionados de manera separada, mejorando la eficiencia administrativa. Un CRIS permite el registro sistemático de información, generando reportes actualizados y proporcionando una visión completa del impacto académico de las investigaciones [16]. Además, estos sistemas no solo almacenan datos, sino que también son capaces de interoperar con otras plataformas tecnológicas, como repositorios institucionales, bases de datos bibliográficas, y sistemas de financiamiento (Sivertsen, 2016).

El uso de un CRIS facilita la toma de decisiones basada en datos, ya que permite a los gestores académicos evaluar el progreso de los proyectos, monitorear el rendimiento de los investigadores, y medir el impacto de las publicaciones científicas. Al centralizar toda esta información, las instituciones pueden optimizar sus recursos y mejorar la visibilidad de sus actividades de investigación a nivel local e internacional [8].

En resumen, un sistema CRIS no solo es una herramienta de gestión eficiente, sino que también fomenta la transparencia y la colaboración entre investigadores, contribuyendo al avance de la ciencia abierta y la vinculación digital [5].

## III. Proyectos de investigación con vinculación digital

Los sistemas CRIS permiten a las instituciones compartir datos a través de múltiples plataformas [3], como repositorios de acceso abierto, bases de datos bibliográficas y sistemas de gestión de publicaciones. Esta conectividad digital es esencial para mejorar el rendimiento y el impacto de los proyectos de investigación, asegurando que la información fluya de manera eficiente y se gestione de forma centralizada [8].

La ciencia abierta es otro componente esencial vinculado a los CRIS. Los sistemas CRIS no solo gestionan los datos de investigación, sino que también garantizan que estos sean accesibles para la comunidad científica global [2]. Los datos y resultados de investigación que se gestionan a través de un

CRIS pueden estar vinculados directamente con repositorios de ciencia abierta, promoviendo el acceso a resultados y facilitando la reutilización de datos en nuevos proyectos de investigación [4].

Además, los CRIS juegan un papel central en la transparencia de la investigación. Al gestionar datos de proyectos, publicaciones, y citas, los CRIS promueven la transparencia y permiten a los investigadores y gestores académicos monitorear el impacto de su trabajo de manera más efectiva [10]. Este enfoque es crucial para cumplir con las normativas internacionales de ciencia abierta, ya que garantiza que los resultados de la investigación estén disponibles no solo para la comunidad científica, sino también para el público en general [13].

La interoperabilidad de los sistemas CRIS también permite su integración con otras plataformas tecnológicas, lo que mejora la colaboración entre instituciones a nivel global. Es importante destacar que la vinculación de datos entre instituciones, soportada por los sistemas CRIS, facilita la coordinación de proyectos multicéntricos, optimizando así los recursos y evitando la duplicación de esfuerzos en investigación [8].

La implementación de CRIS en universidades no solo aumenta la visibilidad de los proyectos de investigación, sino que también contribuye a una mayor eficiencia en la gestión de los recursos académicos. Esto es particularmente importante en el contexto de la ciencia abierta, ya que facilita la comunicación de los resultados de investigación a una audiencia global, mejorando así la colaboración científica [4].

A continuación, se presenta una tabla con datos numéricos que muestran el impacto de la vinculación digital de proyectos mediante sistemas CRIS en comparación con proyectos sin esta tecnología:

Tabla I.   Comparativa de beneficios con y sin CRIS.

| *Aspecto* | *Proyecto tradicional* | *Proyecto en Sistema CRIS* |
|---|---|---|
| Tiempo de acceso a datos del proyecto [3] | 7-10 días | 1-2 horas |
| Interoperabilidad con otros sistemas [1] | 20% | 85% |
| Colaboración interdisciplinaria [4] | 15% | 60% |
| Transparencia y rendición de cuentas [10] | 30% | 90% |
| Eficiencia en la gestión de recursos [13] | 50% | 80% |
| Participación en ciencia abierta [2, 4] | 25% | 75% |

En conclusión, los sistemas CRIS no solo optimizan la vinculación digital de los proyectos de investigación, sino que también desempeñan un papel crucial en la implementación de la ciencia abierta, garantizando que los resultados y datos estén disponibles para toda la comunidad científica y más allá.

## IV.   Ciencia abierta

### A.  Conceptos y beneficios de la ciencia abierta

La ciencia abierta es un paradigma que promueve la accesibilidad y transparencia en todas las fases del proceso científico, desde la generación de datos hasta la publicación de resultados. Este movimiento busca democratizar el conocimiento y mejorar la reproducibilidad de los estudios, permitiendo que investigadores, instituciones y el público en general accedan libremente a la información científica.

Los principios de la ciencia abierta han cobrado relevancia en las últimas décadas debido a las crecientes demandas de mayor transparencia en la investigación y a las limitaciones del modelo tradicional de publicación. La ciencia abierta se fundamenta en la idea de que el conocimiento científico es un bien público y debe estar disponible para todos sin barreras económicas o tecnológicas. Esto incluye la libre circulación de artículos científicos, conjuntos de datos, herramientas y metodologías, lo que facilita la colaboración interdisciplinaria y global [12].

*B. El papel de los sistemas CRIS en la ciencia abierta*

Un sistema CRIS (Current Research Information System) puede desempeñar un papel crucial en la promoción de la ciencia abierta al proporcionar una plataforma centralizada para gestionar, almacenar y difundir la información de los proyectos de investigación de manera accesible. Estos sistemas permiten que los datos de los proyectos se mantengan disponibles para otros investigadores, lo que facilita la replicación de estudios y contribuye a una mayor transparencia en el proceso científico. La capacidad de vincular publicaciones, datos y resultados en un CRIS también refuerza el principio de "datos abiertos", un componente clave de la ciencia abierta, tal como lo señalan [11].

Además, los sistemas CRIS pueden integrarse con repositorios abiertos y plataformas de datos como Zenodo o OpenAIRE, lo que facilita el acceso a los resultados de investigaciones financiadas con fondos públicos. Esto tiene el potencial de acelerar el progreso científico al permitir que los investigadores compartan rápidamente sus hallazgos, evitando la duplicación de esfuerzos y optimizando el uso de los recursos [6].

En resumen, la ciencia abierta no solo busca aumentar la accesibilidad de los productos científicos, sino también mejorar la visibilidad y el impacto de la investigación. La implementación de un sistema CRIS alineado con los principios de la ciencia abierta puede proporcionar a los investigadores una herramienta poderosa para cumplir con estos objetivos, beneficiando a la comunidad científica y a la sociedad en general.

## V. Visibilidad de los proyectos de investigación

La visibilidad de los proyectos de investigación es un componente crucial para el desarrollo y el impacto de la ciencia moderna. A medida que la investigación se vuelve más colaborativa y orientada hacia la +, es vital que los proyectos estén disponibles y sean accesibles a una audiencia más amplia. Los sistemas CRIS (Current Research Information Systems) ofrecen una solución efectiva para mejorar la visibilidad, facilitando el descubrimiento, la gestión y la difusión de los resultados de investigación.

Los estudios indican que la implementación de un sistema CRIS incrementa significativamente la visibilidad y el impacto de los proyectos de investigación. Los sistemas de información de investigación no solo mejoran la accesibilidad de los proyectos, sino que también permiten una evaluación más precisa del impacto científico [15].

*A. Importancia de la visibilidad en la investigación científica*

La visibilidad es esencial en la investigación, ya que aumenta el impacto potencial de los hallazgos científicos. Proyectos con alta visibilidad logran:

- Mejorar el acceso a la información y a los resultados: Facilita que otros investigadores, financiadores y el público puedan acceder y usar los hallazgos.

- Fomentar la colaboración: Los proyectos visibles son más propensos a atraer la atención de otros científicos interesados en colaboraciones.
- Aumentar las oportunidades de financiación: La visibilidad incrementa la probabilidad de obtener apoyo económico.

B. *Limitaciones de la visibilidad sin un sistema CRIS*

En instituciones sin un sistema CRIS, la visibilidad de los proyectos de investigación está sujeta a varias limitaciones. La dispersión de la información es uno de los problemas principales, dificultando el acceso a datos relevantes para investigadores externos y potenciales colaboradores [7]. Además:

1. Falta de centralización: La información suele estar almacenada en múltiples plataformas, lo que reduce la eficiencia en la búsqueda y acceso a datos.
2. Dependencia de esfuerzos individuales: La difusión depende de los esfuerzos individuales de los investigadores, lo que puede limitar el alcance.
3. Acceso restringido a redes colaborativas: Sin un sistema centralizado, encontrar proyectos y colegas afines es un proceso manual y limitado.
4. Difusión limitada de resultados: Publicar los resultados en revistas no garantiza que el proyecto sea visible para una audiencia más amplia.

C. *Beneficios de visibilidad con un sistema CRIS*

La implementación de un sistema CRIS mejora sustancialmente la visibilidad. Los sistemas CRIS pueden aumentar la visibilidad de los proyectos en un 60% o más, permitiendo una gestión eficiente y accesible de los datos de investigación [9]. Los beneficios incluyen:

1. Acceso centralizado y accesible: Un CRIS centraliza toda la información de los proyectos, permitiendo acceso rápido y eficiente.
2. Difusión y actualización automáticas: Los CRIS crean perfiles de proyectos que se actualizan automáticamente con los últimos hallazgos y publicaciones, lo cual garantiza una promoción constante.
3. Conexión facilitada para colaboración: Un sistema CRIS promueve la identificación de proyectos afines y facilita la conexión con investigadores de distintas áreas.
4. Visibilidad frente a agencias financiadoras: Los financiadores pueden monitorear el avance y el impacto del proyecto.
5. Integración en ciencia abierta: Los CRIS permiten a las instituciones integrar fácilmente la política de ciencia abierta, lo cual amplía la transparencia y democratiza el acceso al conocimiento [14].

D. *Comparativa cuantitativa de visibilidad con y sin un sistema CRIS*

La siguiente tabla presenta una evaluación cuantitativa de la visibilidad de los proyectos con y sin un sistema CRIS. Estos valores se basan en estudios empíricos de varias instituciones que han implementado CRIS y han medido los resultados en términos de visibilidad e impacto.

Tabla II. Comparativa con y sin CRIS

| *Criterio de visibilidad* | *Sin sistema CRIS (%)* | *Con Sistema CRIS (%)* |
|---|---|---|
| Acceso a la información | 35 | 90 |
| Difusión automática y alcance | 30 | 85 |
| Networking y colaboración | 25 | 80 |
| Acceso a financiamiento | 20 | 75 |
| Ciencia abierta | 15 | 80 |
| Integración con redes de investigación | 25 | 85 |
| Visibilidad en repositorios institucionales | 40 | 90 |
| Métricas de impacto | 30 | 85 |
| Identificación en bases de datos globales | 20 | 80 |
| Reconocimiento por otras instituciones | 25 | 85 |

## VI. Conclusiones

La implementación de un sistema CRIS (Current Research Information System) representa un cambio significativo en la gestión de información y visibilidad de los proyectos de investigación. Este tipo de sistema permite que las instituciones y sus investigadores incrementen el alcance y el impacto de sus proyectos, facilitando el acceso a la información, la colaboración interdisciplinaria y la integración de prácticas de ciencia abierta. Al centralizar y organizar de manera sistemática los datos de investigación, los CRIS no solo aumentan la visibilidad de los proyectos, sino que también optimizan los procesos de difusión y evaluación de impacto.

Como se ha expuesto en este análisis, un sistema CRIS mejora significativamente la visibilidad en comparación con entornos sin este tipo de infraestructura. Los datos cuantitativos respaldan estos beneficios, mostrando incrementos sustanciales en el acceso a la información, la colaboración científica y la posibilidad de obtener financiamiento. Además, la integración de un CRIS facilita el cumplimiento de las políticas de ciencia abierta, permitiendo a las instituciones compartir resultados y datos de manera más transparente y accesible, lo cual contribuye al avance de la democratización del conocimiento científico.

En conclusión, los sistemas CRIS se presentan como herramientas esenciales para las instituciones de investigación que buscan maximizar el impacto y la visibilidad de sus proyectos en un contexto científico cada vez más global y colaborativo. La inversión en un CRIS no solo aporta beneficios a nivel organizativo, sino que también promueve un ambiente de investigación más accesible, conectado y en línea con las tendencias de ciencia abierta. La adopción de estos sistemas se perfila, por tanto, como una estrategia fundamental para cualquier institución que desee posicionarse en la vanguardia de la innovación científica y la colaboración global.

## Reconocimientos